\documentclass[sigconf,natbib=false,authorversion]{acmart}

\AtBeginDocument{%
  }

\usepackage{booktabs}
\usepackage{siunitx}
\usepackage{color}
\usepackage{transparent}
\usepackage{svg}
\usepackage{multirow}
\usepackage{tikz}

\usepackage[english]{babel}
\usepackage{blindtext}
\usepackage{csquotes}
\usepackage{hyperref}

\usepackage{tabularx}
\usepackage{booktabs}
\usepackage{subcaption}
\usepackage{multicol}

\usepackage{pgfplots}
\usepgfplotslibrary{groupplots,dateplot}
\usetikzlibrary{patterns,shapes.arrows}
\pgfplotsset{compat=newest}

\usepgfplotslibrary{external}
\def\clap#1{\hbox to 0pt{\hss#1\hss}}

\usepackage{xspace}
\newcommand*{\eg}{e.g.\@\xspace}
\newcommand*{\ie}{i.e.\@\xspace}

\usepackage{tikz}
\usetikzlibrary{positioning}
\usetikzlibrary{scopes}
\usetikzlibrary{calc}
\usetikzlibrary{decorations.pathmorphing}
\usetikzlibrary{decorations.pathreplacing}
\usetikzlibrary[shapes.symbols]
\usetikzlibrary[backgrounds]
\usetikzlibrary[fit]
\usetikzlibrary{matrix}

\definecolor{mplblue}{RGB}{31,119,180}

\usepackage{xparse}
\makeatletter
\NewDocumentCommand {\getnodedimen} {O{\nodewidth} O{\nodeheight} m} {
  \begin{pgfinterruptboundingbox}
  \begin{scope}[local bounding box=bb@temp]
    \node[inner sep=0pt, fit=(#3)] {};
  \end{scope}
  \path ($(bb@temp.north east)-(bb@temp.south west)$);
  \end{pgfinterruptboundingbox}
  \pgfgetlastxy{#1}{#2}
}
\makeatother

\def\Pfam{P_{\mathrm{fam}}}%
\def\Pnew{P_{\mathrm{new}}}%
\def\Psameas{P_{\mathrm{same\_AS}}}%

\copyrightyear{2023}
\acmYear{2023}
\setcopyright{acmlicensed}\acmConference[ARES 2023]{The 18th International
Conference on Availability, Reliability and Security}{August 29-September 1,
2023}{Benevento, Italy}
\acmBooktitle{The 18th International Conference on Availability, Reliability and
Security (ARES 2023), August 29-September 1, 2023, Benevento, Italy}
\acmPrice{15.00}
\acmDOI{10.1145/3600160.3600164}
\acmISBN{979-8-4007-0772-8/23/08}

\RequirePackage[
  datamodel=acmdatamodel,
  style=acmnumeric,
  mincrossrefs=1000,
  ]{biblatex}

\begin{document}

\title{Modeling Tor Network Growth by Extrapolating Consensus Data}

\author{Christoph Döpmann}
\email{christoph.doepmann@tu-berlin.de}
\orcid{0000-0001-9689-259X}
\affiliation{%
  \institution{Technische Universität Berlin}
  \department{Distributed Security Infrastructures}
  \city{Berlin}
  \country{Germany}
}
\author{Florian Tschorsch}
\email{florian.tschorsch@tu-berlin.de}
\orcid{0000-0001-9689-259X}
\affiliation{%
  \institution{Technische Universität Berlin}
  \department{Distributed Security Infrastructures}
  \city{Berlin}
  \country{Germany}
}

\begin{abstract}
Since the Tor network is evolving into an infrastructure for anonymous communication,
analyzing the consequences of network growth is becoming more relevant than ever.
In particular, adding large amounts of resources may have unintentional consequences
for the system performance as well as security.
To this end, we contribute a methodology for the analysis of scaled Tor networks
that enables researchers to leverage real-world network data.
Based on historical network snapshots (consensuses),
we derive and implement a model for methodically scaling Tor consensuses.
This allows researchers to apply established research methods to scaled networks.
We validate our model based on historical data, showing its applicability.
Furthermore, we demonstrate the merits of our data-driven approach
by conducting a simulation study to identify performance impacts of scaling Tor.
\end{abstract}



\maketitle

\section{Introduction}

Over the years, the Tor network~\cite{dingledine2004tor} has experienced considerable growth,
representing the increasing need for online anonymity.
Apart from this general trend, however,
several developments indicate that
Tor will (have~to) grow even much further over the upcoming years.
Firstly, with the introduction of the new official Tor client \emph{arti}~\cite{arti-announcement},
Tor opens to a much broader user base that goes beyond its current main usage in a web browser.
Secondly, more web browsers consider tightly integrating Tor support
as part of their private browsing modes.~\cite{brave-using-tor,firefox-spbm}.
As a consequence, there is a clear need for a well-founded strategic plan
on how to accommodate such growth in the Tor network,
particularly how to scale the network.

\begin{figure}[t]
    \centering
    {
        \tikzset{external/export=false}
        \input{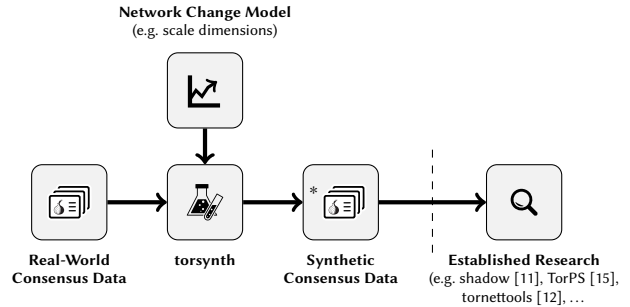}
    }
    \caption{
        Proposed research workflow.
        Real-world Tor consensus data is processed and modified
        to enable research on extrapolated networks.
    } \label{fig:system-overview}
\end{figure}

When adding resources to the Tor network,
there are two general dimensions of growth that are
in line with prior research on distributed systems~\cite{DBLP:conf/ipps/MichaelMSW07}:
horizontal growth (adding new relays) and
vertical growth (increasing the existing relays' bandwidth).
Today, however, there is no method that can guide the decision-making process
on the question of which direction would be preferable,
and what would be the implications for the Tor network.
While relays are generally operated by third parties
and are not subject to control by the Tor project,
knowing the implications of either growth dimension
would allow to identify upcoming issues early.
Also, this may allow to incentivize the preferable kind of resource donation,
steering the network into one or the other direction.
These goals, however, require thorough insight
on the implications of the two scaling dimensions,
affecting performance as well as security.

Over the years, the large body of Tor research~\cite{DBLP:journals/csur/AlSabahG16}
established a series of research methods.
One of the most prominent approaches uses real-world snapshots of the network,
the so-called \emph{consensus}, for analyzing Tor.
This has resulted in a variety
of tooling and methodology~\cite{denasa,point-break,users-get-routed,waterfilling,tempest}.
As of today, however, these data-driven approaches can only be applied
for the analysis of the network \emph{history},
which is not sufficient to understand and tackle upcoming challenges.

In this paper, we aim to fill this methodical gap.
As our main contribution, we provide a methodology to
extrapolate real-world Tor consensus documents into scaled versions of the Tor network.
We develop models for horizontal and vertical growth of the Tor network
that serve as input to \emph{torsynth},
a software framework to generate synthetic consensus documents.
This allows us and fellow researchers to apply established research methodologies%
---for instance, network simulation and circuit generation analysis.
In Figure~\ref{fig:system-overview}, we illustrate this data pipeline
and how our work will benefit researchers working in this field.
Please note, we do \emph{not} aim to predict network growth in Tor
but instead establish the foundations for exploring and experimenting with
different developments that may be conceivable,
which can also include predictive models in the future.

As our first step, we motivate our focus on horizontal and vertical network growth
by analyzing the past growth behavior of the Tor network,
based on historical data~(Sec.~\ref{sec:history}).
We then introduce a novel network growth model
as well as its implementation~(Sec.~\ref{sec:model}),
and validate its applicability based on historical data~(Sec.~\ref{sec:model-validation}).
Finally, we demonstrate its utility
by carrying out large-scale simulations~(Sec.~\ref{sec:eval})
of \emph{extrapolated versions} of the Tor network
using the shadow simulator~\cite{jansen2012shadow}.
We analyze the performance implications of different scaling dimensions,
coming to the non-obvious finding
that focusing only on horizontal growth instead of vertical scaling
may harm the performance of the Tor network.

\section{Evolution of the Tor Network} \label{sec:history}

In order to motivate our model for extrapolating Tor network consensuses,
we examine and characterize Tor's evolution of network growth in the past ten~years.

\subsection{Tor Basics}

Tor achieves anonymity for its users based on
the principle of \emph{onion routing}~\cite{goldschlag1996onionrouting}.
That is, each user's stream of communication is transferred
over a path of intermediate \emph{relays},
called a \emph{circuit}.
Relay information is collected and published by \emph{directory authorities}
who compile their view of the network into \emph{consensus documents}.
These are used by clients to select relays for their circuits.
This choice is made randomly but weighted by the relays' \emph{bandwidth},
which is measured for validation by \emph{bandwidth authorities}.
Relays can be used for different positions in a circuit,
depending on the \emph{flags} they have been assigned (\eg, exit or guard relays).
For relays, the so-called family relation can be set by the operator.
If two relays are mutually defined as family members,
they are understood to be operated by the same person or organization
and are not chosen together in a circuit.
We refer the reader to the specification~\cite{dirspec} for more in-depth information.

\subsection{Tor Network Growth} \label{sec:growth}

By analyzing the Tor network size from the past,
we generate first insights on the dimensions of network growth
as a baseline for building our model.
In particular, we focus on validating
whether the general differentiation into horizontal and vertical growth,
known from other distributed systems,
also applies specifically to Tor.
Horizontal network growth is due to new relays joining the network.
On the other hand, vertical growth refers to the \enquote{size} of the individual relays,
\ie, their bandwidth characteristics.

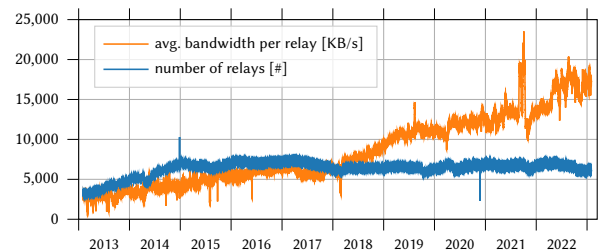
\begin{figure}[b]
    \centering
        \pgfplotsset{width=\columnwidth, height=0.5\linewidth}
        \pgfplotsset{
            tick label style={font=\scriptsize\sffamily},
            scaled ticks=false,
            label style={font=\scriptsize\sffamily},
            y label style={yshift = -0.01\textwidth},
            legend style={font=\scriptsize\sffamily},
        }
        \begin{tikzpicture}

\definecolor{color0}{rgb}{0.12156862745098,0.466666666666667,0.705882352941177}
\definecolor{color1}{rgb}{1,0.498039215686275,0.0549019607843137}

\begin{axis}[
date coordinates in=x,
legend pos={north west},
legend cell align={left},
legend style={fill opacity=0.8, draw opacity=1, text opacity=1, draw=white!80!black},
scaled x ticks=manual:{}{\pgfmathparse{#1}},
tick align=outside,
x grid style={white!69.0196078431373!black},
xmajorgrids,
xmin=2013-01-01 00:00, xmax=2023-03-15 00:00,
xtick style={color=black},
xtick={
    2013-01-01 00:00,
    2014-01-01 00:00,
    2015-01-01 00:00,
    2016-01-01 00:00,
    2017-01-01 00:00,
    2018-01-01 00:00,
    2019-01-01 00:00,
    2020-01-01 00:00,
    2021-01-01 00:00,
    2022-01-01 00:00,
    2023-01-01 00:00
},
xticklabels={2013,2014,2015,2016,2017,2018,2019,2020,2021,2022},
x tick label as interval,
y grid style={white!69.0196078431373!black},
ymajorgrids,
ymin=0, ymax=25000,
ytick pos=left,
ytick style={color=black},
ytick={0,5000,10000,15000,20000,25000},
]
\addplot [semithick, color1]
table [header=false,col sep=comma] {%
figures/history-raw/history-raw-001.tsv};
\addlegendentry{avg. bandwidth per relay [KB/s]}
\addplot [semithick, color0]
table [header=false,col sep=comma] {%
figures/history-raw/history-raw-000.tsv};
\addlegendentry{number of relays [\#]}
\end{axis}

\end{tikzpicture}
        \caption{Size of the Tor network over the past 10~years.}
        \label{fig:history-raw}
\end{figure}
\begin{figure}[t]
        \pgfplotsset{width=\columnwidth, height=0.5\linewidth}
        \pgfplotsset{
            tick label style={font=\scriptsize\sffamily},
            scaled ticks=false,
            label style={font=\scriptsize\sffamily},
            y label style={yshift = -0.01\textwidth},
            legend style={font=\scriptsize\sffamily},
        }
        \begin{tikzpicture}

\definecolor{color0}{rgb}{0.12156862745098,0.466666666666667,0.705882352941177}
\definecolor{color1}{rgb}{1,0.498039215686275,0.0549019607843137}

\begin{axis}[
date coordinates in=x,
legend cell align={left},
legend style={fill opacity=0.8, draw opacity=1, text opacity=1, draw=white!80!black},
tick align=outside,
tick pos=left,
x grid style={white!69.0196078431373!black},
xmajorgrids,
xmin=2013-01-01 00:00, xmax=2023-03-15 00:00,
xtick style={color=black},
xtick={
    2013-01-01 00:00,
    2014-01-01 00:00,
    2015-01-01 00:00,
    2016-01-01 00:00,
    2017-01-01 00:00,
    2018-01-01 00:00,
    2019-01-01 00:00,
    2020-01-01 00:00,
    2021-01-01 00:00,
    2022-01-01 00:00,
    2023-01-01 00:00
},
xticklabels={2013,2014,2015,2016,2017,2018,2019,2020,2021,2022},
x tick label as interval,
y grid style={white!69.0196078431373!black},
ymajorgrids,
ymin=0.5, ymax=2.5,
ytick distance=0.25,
ytick style={color=black}
]

\addplot[color=black,thick, forget plot] coordinates {(2012-01-01 00:00,1) (2023-01-01 00:00,1)};

\addplot [semithick, color0, opacity=0.2, forget plot]
table [header=false,col sep=comma] {%
figures/history-analysis/history-analysis-000.tsv};
\addplot [semithick, color1, opacity=0.2, forget plot]
table [header=false,col sep=comma] {%
figures/history-analysis/history-analysis-002.tsv};
\addplot [semithick, color0]
table [header=false,col sep=comma] {%
figures/history-analysis/history-analysis-001.tsv};
\addlegendentry{annual horizontal growth $H(t)$}
\addplot [semithick, color1]
table [header=false,col sep=comma] {%
figures/history-analysis/history-analysis-003.tsv}; 
\addlegendentry{annual vertical growth $V(t)$}
\end{axis}

\end{tikzpicture}
        \caption{Computed vertical and horizontal growth rate of the Tor network
                 over the past 10~years: raw values (shaded), and 90-days moving average (solid).}
        \label{fig:history-analysis}
\end{figure}
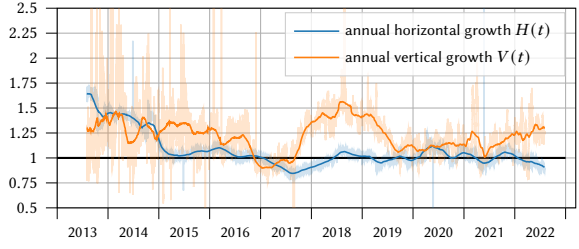

We base our analysis on historical Tor consensus data from~\cite{tor-metrics},
covering the past 10~years (2013-02 to 2023-01).
We make use of the following base metrics upon which we build our growth rate indicators.
For any given time~$t$, we define $n(t)$ as the number of relays at time~$t$.
We use this to estimate horizontal growth.
Moreover, we define $b(t)$ as the average bandwidth per relay at that time
and use this as an indicator for vertical growth.

These base metrics
are evaluated in Figure~\ref{fig:history-raw}.
On first sight, one can see that these two network dimensions
have indeed developed differently over time.
While, in the beginning of the timeframe,
both horizontal and vertical network growth are present
(and especially a growth in the number of relays can be observed),
this coupling has later come to an end.
Especially in~2018 and early~2019,
network growth was primarily due to an increasing average bandwidth per relay.
On close investigation, we can see that the growth dynamics of the network
have evolved in relatively distinct phases.

In order to quantify this behavior more rigorously, we took the following approach.
For each point in time~$t$ and a to-be-defined growth time span~$\Delta T$,
we define the horizontal growth $H(t)$ and the vertical growth~$V(t)$ as follows:

\[
H(t) = \frac{n \left( t + \frac{\Delta T}{2} \right)}{n \left( t - \frac{\Delta T}{2} \right)}, \qquad
V(t) = \frac{b \left( t + \frac{\Delta T}{2} \right)}{b \left( t - \frac{\Delta T}{2} \right)}
\]

This way, we understand the network growth rate at time~$t$
as the factor by which the network has grown in the time span~$\Delta T$ around~$t$.
In order to obtain readable results, we set $\Delta T = 1~\mathrm{year}$.
Consequently, the resulting values for $H(t)$ and $V(t)$ can be interpreted
as the annual growth rate, interpolated down to a single point $t$ in time.
A rate of $1.0$ means that the network did neither grow nor shrink.
For visual clarity, we also apply a 90-days moving average on top of $H(t)$ and $V(t)$.

Figure~\ref{fig:history-analysis} displays the results.
The computed growth rates support our initial, visual impression.
In the first years of the time span,
there was both horizontal and vertical growth.
Horizontal growth has mainly came to a cease around~2015.
Since then, also vertical growth has declined,
with the notable exception of the aforementioned period around~2018,
in which the network reached an annual vertical growth of up~to~150\%.
Today, there is a steady but small amount of vertical growth,
while the number of relays has not significantly grown again.
While we can only speculate about the exact reasons,
we note that vertical growth also comes from performance improvements
that affect all relays simultaneously.
One example are runtime improvements that stem from code optimization.
Therefore, ongoing efficiency improvements of the Tor~client
implicitly lead to vertical growth.

Our data analysis shows that one can clearly distinguish
historical phases in which the Tor network has grown in different dimensions%
---horizontal and vertical.
Both are relevant and have significantly shaped the network in the past.
We therefore regard both dimensions as valid and significant factors
that are relevant for understanding Tor growth dynamics
as well as the associated implications.
Thus, they will serve as the basis for our network growth model.

\section{Scaling Tor Consensuses} \label{sec:model}

We now focus on the question how to generate synthetic Tor network consensuses
that capture a larger Tor network.
Since it is generally unknown how the Tor network will grow in the future,
we do \emph{not} intend to \emph{predict} this growth,
but scale an \emph{existing} consensus in a representative way.
We thus give researchers the ability
to explore the impact of different hypothetical scenarios.
Our focus is on the distinction between vertical and horizontal growth,
as motivated in Section~\ref{sec:history}.
In reality, the network will likely see a mixture of these two growth dimensions,
as it has been in the past. %
While our model can be used to apply such mixed growth patterns,
investigating the distinct dimensions is still relevant,
because it allows us to generate insight and build up an understanding
of the benefits and drawbacks of either approach.
Growing the network in different dimensions may have considerable implications
on aspects like performance, anonymity guarantees, diversity, and censorship resistance.
In the following, we present how to model vertical and horizontal scaling.

In order to introduce our approach formally, we rely on the following notation:
By $r_i$, we denote relays in a consensus~$C$, which is defined as~$C = \{r_i\}$.
Let $b_i$ be the bandwidth of relay~$r_i$.
If we create a new relay~$r'$ utilizing some relay~$r$ as template
but with modified bandwidth, then we denote this relay~$r' = r[b=\dots]$.
Furthermore, we make the following two simplifying assumptions:
First, since we focus on the consensus, we consider the consensus weight only
and use it interchangeably with the terms bandwidth, bandwidth weight, and capacity.
In reality, these values might differ,
because some are provided by the relay operator and some are validated by bandwidth authorities.
Second, we consider the relay \emph{families relation} to be transitive.
That is, we consider relay families as connected components
in a graph of family relations.
If relays $A$ and $B$ are defined to be related in terms of a family, as well as $B$ and $C$,
then we assume $A$ and $C$ to be related as well.
We do this as a means of simplification and because it reflects
the families' meaning of being operated by the same entity.
Another conceivable approach would be to regard only cliques as families,
which however would yield a less strict view on which relays
are controlled by the same real-world entity.

\subsection{Vertical Scaling}

When scaling vertically, only the bandwidth of the individual relays grows,
but not their number.
Modeling this type of growth is relatively straightforward.
We can capture the general case of vertical growth
by defining the new consensus $C'$ as follows:

\[
C' = \{ r_i[b=\sigma(b_i)] \mid r_i \in C \}.
\]
In this definition, $\sigma$ is an arbitrary function that defines the new bandwidth for each relay.
In a very simple scenario, $\sigma$ could just apply a static factor~$f$
to all relay bandwidths:
$\sigma_{\text{uniform}}(r_i) = b_i \cdot f$.
However, it is unlikely that the network will grow completely uniformly.
Instead, considering non-uniform bandwidth distributions for relays
is necessary to analyze the impact of a changing shape of the network,
influencing how much load each single relay experiences.
Researchers should be able to investigate
the implications of different growth assumptions,
not only of an unaltered continuation of the past trends.

We contribute two different $\sigma$~distributions.
First, the scale may depend on the bandwidth quantile of each relay.
This may yield interesting results as relay operators of different relay sizes
may be differently ambitious to grow further.
For example, it is not clear what the implications would be
if only the top relays grew.
Our model accepts any list of factors
for scaling the respective relays based on their bandwidth rank.

As a more flexible model, we allow relays to be scaled according to their flags,
\ie, whether they are in the exit, guard, or middle role.
For this, the model accepts three different scale factors:
$f_{\text{middle}}$, $f_{\text{guard}}$, and $f_{\text{exit}}$.
The provided values, however, cannot be used as-is.
This is due to the fact that relays can be exits and guards \emph{at the same time}.
The factors as used for the $\sigma$~distribution are applied to the single relays.
However, we want to give researchers more intuitive control over the influence of a relay's \emph{role}.
Therefore, the provided values should apply
to the whole group of relays operating in the specific relay role.
We therefore derive per-relay scale factors as follows.
We need to balance between exit-only~(E), guard-only~(G), and guard-and-exit~(D) relays
in a way that realizes the requested group factors.
\def\frel{\bar{f}}
Consequently, we have to derive group factors~$\frel_{\{E,G,D\}}$
that also incorporate the group of relays that are both exit and guard~(D).
For middle relays~(M), we can use~$\frel_M = f_{\text{middle}}$,
since there are no conflicts with other relays.

Let $w_{\{M,G,E,D\}}$ be the total bandwidth of all relays in the respective group.
Then, every solution must satisfy the following equations:
\begin{align}
f_{\text{guard}} \cdot (w_G + w_D) &= w_G \cdot \frel_G + w_D \cdot \frel_D \\
f_{\text{exit}}  \cdot (w_E + w_D) &= w_E \cdot \frel_E + w_D \cdot \frel_D
\end{align}

This defines a linear equation system, but does not give a unique solution.
In fact, the additional bandwidth could be balanced arbitrarily between the groups E, G, and D.
For example, $\frel_D = 0$, $\frel_E = f_{\text{exit}}$, and $\frel_G = f_{\text{guard}}$,
which would obviously not make much sense because the relays
that are both exit and guard the same time
would effectively be excluded from scaling.
On the other hand, when solving the linear system,
we still have to make sure that all factors are non-negative.
We therefore choose the following solution,
depending on whether $f_{\text{guard}}$ or $f_{\text{exit}}$ is smaller:

\begin{align*}
\frel_D &= \min(f_{\text{guard}}, f_{\text{exit}}) \\
\frel_E &=
    \begin{cases}
    f_{\text{exit}} \cdot \frac{ (w_E + w_D) - f_{\text{guard}} \cdot w_D } { w_E } &\text{if $f_{\text{guard}} \leq f_{\text{exit}}$} \\
    f_{\text{exit}} &\text{else} \\
    \end{cases} \\
\frel_G &=
    \begin{cases}
    f_{\text{guard}} &\text{if $f_{\text{guard}} \leq f_{\text{exit}}$} \\
    f_{\text{guard}} \cdot \frac{ (w_G + w_D) - f_{\text{exit}} \cdot w_D } { w_G } &\text{else} \\
    \end{cases}
\end{align*}

Put differently, we apply the smaller role factor to all the relays
both in that respective group as well as the shared one,
and adjust the relay factor for the remaining group accordingly
to match the desired role factors.
By simple transformation of these terms,
we can see that this will always yield non-negative values.

With that, we can define $\sigma(r_i)$ as follows:
\[
\sigma(r_i) =
\begin{cases}
b_i \cdot \frel_M &\text{if $r_i$ is a middle relay} \\
b_i \cdot \frel_E &\text{if $r_i$ is an exit-only relay} \\
b_i \cdot \frel_G &\text{if $r_i$ is a guard-only relay} \\
b_i \cdot \frel_D &\text{if $r_i$ is an exit and guard relay}
\end{cases}
\]

\subsection{Horizontal Scaling}

We now consider scaling Tor horizontally.
Unlike vertical scaling, we need to \enquote{invent} new relays.
Specifically, growing horizontally to a factor of $f$ means that we need to
synthetically generate $(f-1) \cdot |C|$ new relays.
The challenge is that
these new relays are required to be \enquote{representative} of the original consensus.
For example, if the original consensus exhibits correlations between different relay properties,
these should also be represented by the new relays.
Such properties may include bandwidth, the Autonomous System~(AS), country, or relay family.

As the main approach, we choose to generate new relays
by sampling and modifying existing relays.
The benefit is that the new relays will be representative of the original network,
without the need to capture all (potentially non-obvious) correlations by hand.
For creating $n$~relays, we sample uniformly $n$ times
from the existing consensus to choose a \emph{base relay}. %
The base relay is used as the template for the new relay.
Some properties can be copied, notably, the consensus weight value.
Some relay-specific properties,
including the relay's nickname, fingerprint and IP address
have to be generated from scratch.
While it is trivial to generate new nicknames and fingerprints,
it is not as obvious for the relay's IP address.
When assigning relays a new IP address, %
we sample from the IP address space that belongs to the AS of the base relay.
By doing so, we keep the network peculiarities of the base relay.
Creating new IP addresses in a way that is representative for the original relay is relevant,
because circuit construction rules exclude pairs of relays
that share the same /16 subnet, for example.
Our approach mirrors the semantics of representatively scaling the state of the network.
It does not immediately \enquote{predict} the future network
which might require to also incorporate ASes that have not been used before.

One aspect that requires more attention are relay families.
This is important because relay properties may be highly correlated within the family,
\eg, being located in the same country or~AS.
To accommodate family relations,
while keeping the original properties representative of the synthetic consensus,
we proceed as follows.
First of all, let $\Pfam$ be the ratio of relays that are in any family (in the original consensus).
We use this probability for deciding whether a newly generated relay shall be part of a family.
Secondly, we calculate the empirical probability~$\Psameas$ that
if two relays are in the same family, they are also in the same AS.
Moreover, we require a probability $\Pnew$
upon which we will decide whether a new relay that shall be part of a family,
will either join an existing one or be part of the new family.
We leave this value as a configurable parameter
because it solely depends on assumptions on how relay families will grow
(either acquiring more relays, or more families being created).
As such, researchers can incorporate their assumptions
and analyze different hypotheses.
Based on these probabilities, we sample the family properties for new relays.
When joining an existing family, we choose it
by sampling another reference relay (taking into account $\Psameas$)
that is only used for determining the target family.
For new families, we sample their size
based on the original distribution of family sizes.

Similarly to vertical scaling,
we also allow to define different weights for different relay roles
(middle, exit, and guard).
This could be interesting for analysis
because the priorities for relay operators to run, \eg, exit relays, may change over time.
At the same time, correlations like exit-relay-friendly ASes should be preserved.
We therefore offer to specify different weights per relay role.
These are taken into account when sampling new relays, leading to the desired network skew.
However, we have the same issue as for vertical scaling:
Some relays are guard and exit the same time.
We therefore apply the same principle as mentioned above.
However, we now use the \emph{number of relays} per relay group,
instead of their total bandwidth,
and use the resulting factors to weigh the relays during sampling.

\begin{figure*}[t]
        {
            \pgfplotsset{width=.33\linewidth, height=0.20\linewidth}
\pgfplotsset{
    tick label style={font=\scriptsize\sffamily},
    scaled ticks=false,
    label style={font=\footnotesize\sffamily},
    title style={font=\footnotesize\bfseries\sffamily},
    legend style={font=\footnotesize\sffamily},
}

\begin{tikzpicture}

\definecolor{color0}{rgb}{0.12156862745098,0.466666666666667,0.705882352941177}
\definecolor{color1}{rgb}{1,0.498039215686275,0.0549019607843137}

\begin{axis}[
ylabel={CDF [share of relays]},
xlabel={consensus weight},
legend cell align={left},
legend style={fill opacity=0.8, draw opacity=1, text opacity=1, at={(0.97,0.03)}, anchor=south east, draw=white!80!black},
tick align=outside,
tick pos=left,
title={2013-02 -- 2015-01 (T1)},
x grid style={white!69.0196078431373!black},
xmajorgrids,
xmin=-1000, xmax=55400,
xtick style={color=black},
y grid style={white!69.0196078431373!black},
ymajorgrids,
ymin=-0.049992789154889, ymax=1.04984857225267,
ytick style={color=black},
ytick={-0.2,0,0.2,0.4,0.6,0.8,1,1.2},
yticklabels={−0.2,0.0,0.2,0.4,0.6,0.8,1.0,1.2}
]
\addplot [thick, color0]
table {%
figures/validation-cdfs/cdf-2013-02-01--2015-02-01.csv-000.tsv};
\addlegendentry{real}
\addplot [very thick, color1, dashed]
table {%
figures/validation-cdfs/cdf-2013-02-01--2015-02-01.csv-001.tsv};
\addlegendentry{scaled}
\end{axis}

\end{tikzpicture}
\hspace{1em}
\begin{tikzpicture}

\definecolor{color0}{rgb}{0.12156862745098,0.466666666666667,0.705882352941177}
\definecolor{color1}{rgb}{1,0.498039215686275,0.0549019607843137}

\begin{axis}[
xlabel={consensus weight},
legend cell align={left},
legend style={fill opacity=0.8, draw opacity=1, text opacity=1, at={(0.97,0.03)}, anchor=south east, draw=white!80!black},
tick align=outside,
tick pos=left,
title={2018-01 -- 2020-12 (T2)},
x grid style={white!69.0196078431373!black},
xmajorgrids,
xmin=-1000, xmax=95000,
xtick style={color=black},
y grid style={white!69.0196078431373!black},
ymajorgrids,
ymin=-0.0499926782837897, ymax=1.04984624395958,
ytick style={color=black},
ytick={-0.2,0,0.2,0.4,0.6,0.8,1,1.2},
yticklabels={−0.2,0.0,0.2,0.4,0.6,0.8,1.0,1.2}
]
\addplot [thick, color0]
table {%
figures/validation-cdfs/cdf-2018-01-01--2021-01-01.csv-000.tsv};
\addlegendentry{real}
\addplot [very thick, color1, dashed]
table {%
figures/validation-cdfs/cdf-2018-01-01--2021-01-01.csv-001.tsv};
\addlegendentry{scaled}
\end{axis}

\end{tikzpicture}
\hspace{1em}
\begin{tikzpicture}

\definecolor{color0}{rgb}{0.12156862745098,0.466666666666667,0.705882352941177}
\definecolor{color1}{rgb}{1,0.498039215686275,0.0549019607843137}

\begin{axis}[
xlabel={consensus weight},
legend cell align={left},
legend style={fill opacity=0.8, draw opacity=1, text opacity=1, at={(0.97,0.03)}, anchor=south east, draw=white!80!black},
tick align=outside,
tick pos=left,
title={2019-01 -- 2019-12 (T3)},
x grid style={white!69.0196078431373!black},
xmajorgrids,
xmin=-1000, xmax=91606,
xtick style={color=black},
y grid style={white!69.0196078431373!black},
ymajorgrids,
ymin=-0.0499920785804816, ymax=1.04983365019011,
ytick style={color=black},
ytick={-0.2,0,0.2,0.4,0.6,0.8,1,1.2},
yticklabels={−0.2,0.0,0.2,0.4,0.6,0.8,1.0,1.2}
]
\addplot [thick, color0]
table {%
figures/validation-cdfs/cdf-2019-01-01--2020-01-01.csv-000.tsv};
\addlegendentry{real}
\addplot [very thick, color1, dashed]
table {%
figures/validation-cdfs/cdf-2019-01-01--2020-01-01.csv-001.tsv};
\addlegendentry{scaled}
\end{axis}

\end{tikzpicture}
        }
        \caption{Consensus weight distribution:
                 real-world data vs. scaling old consensuses to the same size
                 (limited to lower~99\%).
                 }
        \label{fig:validation-cdfs}
\end{figure*}
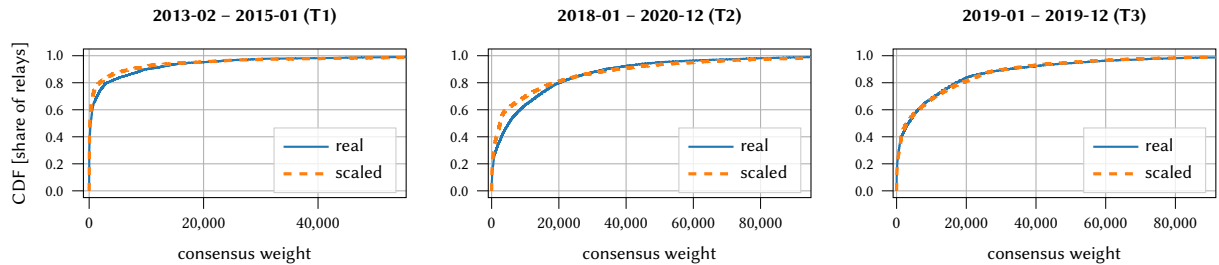

\subsection{Implementation: torsynth}

We implemented the two scaling methods as an initial component
for our open-source framework \emph{torsynth},%
\footnote{\url{https://github.com/cdoepmann/torsynth}}
which is envisioned to become a versatile toolbox
for generating and experimenting with synthetic Tor consensuses.
As input, torsynth takes a consensus document describing
a state of the Tor network,
as can be obtained, \eg, from Tor Metrics~\cite{tor-metrics}.
The extrapolated consensus is output in the same format.
This way, torsynth's output can be fed to existing tools
that were originally intended to analyze
the past evolution of Tor~(cf.~Figure~\ref{fig:system-overview}).

The two scaling dimensions are implemented independently,
but can trivially be combined by being applied consecutively.
Also, the two scaling methods operate on a single consensus only.
While it has been shown that there is large churn in the Tor network
and it is preferable to operate on averaged network snapshots~\cite{once-never-enough},
this is fully compatible because our scaling methods
can operate not only on original Tor consensuses,
but on representative samples as generated, \eg, by tornettools~\cite{once-never-enough}, as well.

\section{Model Validation} \label{sec:model-validation}

In order to build confidence in the validity of our proposed network scaling model,
we carry out an initial validation.
We aim to investigate whether Tor consensuses that have been scaled using \emph{torsynth}
are in fact representative for an equivalent amount
of \enquote{natural} growth of the Tor network.

We do so by relying on historical data as the ground truth to compare against.
Our approach is to have torsynth scale a historical consensus
and compare it to the change that happened in the real world.
More specifically, we choose two consensuses $A$ and $B$ from the past,
where $B$ occurred some time after $A$.
We then compute the amount of horizontal and vertical growth that happened between them.
Given these scale factors, we use torsynth to scale the original consensus $A$,
generating a synthetic consensus $A'$, which we compare against $B$.
If our model mirrors the real growth behavior, the consensuses $A'$ and $B$ should be similar.

Based on our analysis in Section~\ref{sec:growth},
we identify different historical time periods that are of special interest,
and choose our validation periods accordingly.
Firstly, the time from 2013-02 to 2015-01~(T1), a two-year period,
in which both horizontal and vertical growth was high.
Secondly, the three years of~2018 to~2020~(T2),
in which there was mainly vertical growth that slowed down over time.
And thirdly, we selected a one-year sub-period of this (year~2019 only, T3)
to demonstrate behavior on a shorter time span.
We apply the aforementioned scaling procedure
and compare the results primarily based on the CDF of bandwidth weights the consensuses' relays exhibit.

We visualize the results in Figure~\ref{fig:validation-cdfs}.
As can be seen, the synthetically scaled consensuses closely resemble the real-world data.
In particular, the overall resource distributions within the network are of very similar shape.
For readability of the CDF~plots, we have omitted the upper~1\% (long tail).
Thorough analysis reveals, though,
that the few largest relays are the only ones where a clear difference can be found.
There, vertical growth was either overestimated or underestimated by our model.
This effect was to be expected as the extreme values within the network
solely depend on the singular activities of a few relay operators
and thus cannot be captured by an extrapolating, continuous model like ours.
In order to quantify the quality of the scaled consensuses,
we calculate the per-rank bandwidth difference between relays of the two distributions,
scaling it to the average relay consensus weight.
The resulting per-cent value denotes the deviation
of a single relay's bandwidth weight in the scaled consensus from the real-world data.
We obtain the following median values:
3.4\% for~T1, 12.0\% for~T2, and 2.4\% for~T3.
These numbers again underline the ability of our model
to reproduce real-world network growth (T1~and~T3).
At the same time, they show the limitation of purely modelled growth synthesis
if the considered time span is too large.
The model error grows with the length of the considered time span.
As such, during~T2 which spans 3~years,
the Tor network has experienced considerable structural change that is not captured by our model.
Instead, torsynth reproduces the network properties from the beginning of this time period,
as intended.

All in all, we can conclude that torsynth is effective at scaling Tor consensuses
in a way that transfers the present network properties to scaled, synthetic consensuses.

\section{Case Study: Performance Implications of Scaling Tor} \label{sec:eval}

We leverage our methodology of scaling Tor consensuses
for an initial exploration of the implications of scaling the network.
On the one hand, we do so in order to demonstrate the utility
of our methodology for the empirical analysis of extrapolated Tor networks.
On the other hand, we aim to utilize simulations
for generating initial insight on the consequences of growing Tor,
which introduces a novel perspective due to its reliance on extrapolated real-world data.
In particular, we provide a first impression on
how performance is affected by the different scaling dimensions.
To this end, we carry out a series of simulations
of the original as well as scaled Tor networks.
We investigate whether there are significant changes in performance
if Tor is scaled either horizontally or vertically, compared to the original Tor network.
While we focus on performance here,
other aspects like security and anonymity require further investigation
and can also be analyzed using our methodology.
For this  paper, however, they are out of scope.

We carried out network simulations using the shadow simulator~\cite{jansen2012shadow}.
For this, we followed the typical workflow:
We used tornettools~\cite{once-never-enough}
to prepare the scenario for shadow,
instantiating its configuration files appropriately.
By sampling subnetworks from our scaled Tor networks,
tornettools also makes the simulation feasible,
reducing the computational complexity~\cite{once-never-enough}.
It is thus an example on how torsynth integrates seamlessly into existing experimentation workflows.
Based on our model as laid out in Section~\ref{sec:model},
we separately scaled the network to~$f=2.0$ horizontally and vertically,
and compared the resulting network performance metrics obtained from shadow simulations on them.
For the scaled network scenarios, we also increased the number of clients proportionally.
We used tornettools to generate 10\% sample networks from each of the scales networks
as well as the original consensus (which was from 2022-03-17-00-00-00).
For each of the variants (unscaled, vertical, horizontal), we repeated the procedure 10~times.
Each of the, in total, 30~simulation runs took up to 40~hours and used up to 550~GB of memory.
As in~\cite{once-never-enough}, we calculated the appropriate 95\% confidence intervals per quantile
on the collected performance metrics data.

\begin{figure}[t]
    \begin{subfigure}{0.32\textwidth}
        \includegraphics[width=\textwidth]{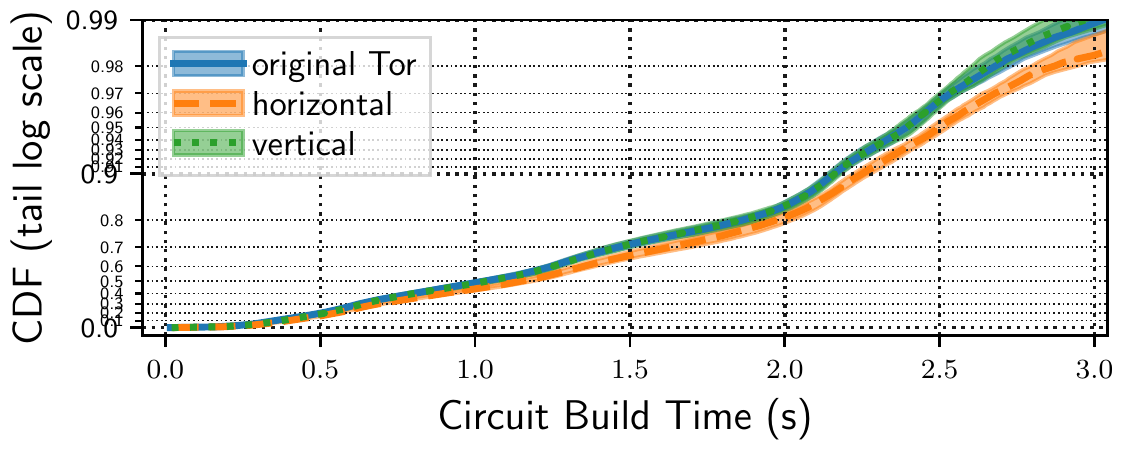}
    \end{subfigure}
    \begin{subfigure}{0.32\textwidth}
        \includegraphics[width=\textwidth]{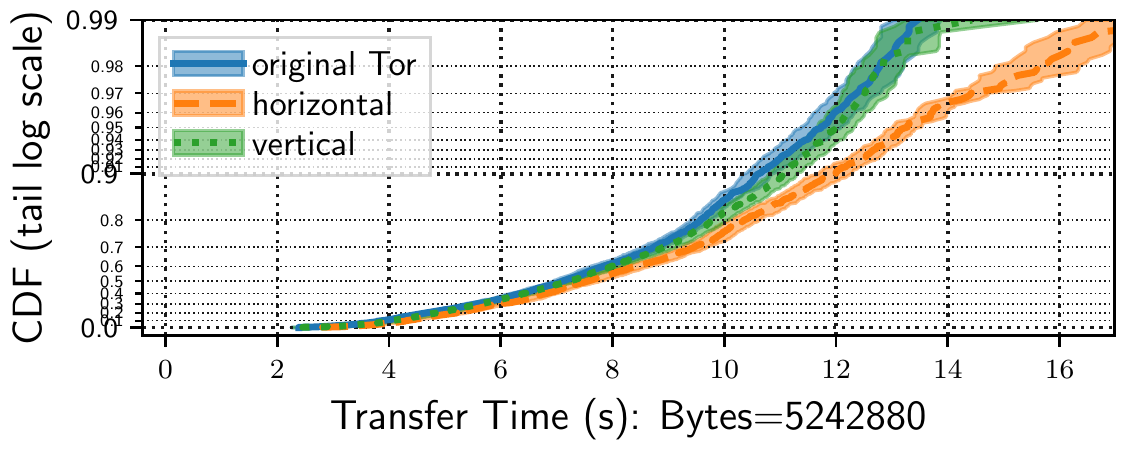}
    \end{subfigure}
    \caption{Performance metrics for Tor clients in shadow simulations of the original Tor network,
             a horizontally scaled version, and a vertically scaled version.
             The shaded areas represent 95\% confidence intervals per quantile.
    } \label{fig:results-horz-vert}
\end{figure}

Figure~\ref{fig:results-horz-vert} presents our results.
The plots display cumulative distribution functions over the circuits of a simulation run,
along with the 95\%~per-quantile confidence interval over the different simulation runs.
We first consider the time it takes to establish a new circuit,
an important metric for the interactive usability of Tor.
It can be noted that the vertically scaled Tor network
behaves very much like the original one.
In contrast, when scaling the network horizontally,
there is a small but significant degradation.
The circuit build times are worse especially in the long tail,
but also for most other circuits.
when looking at the transfer time of 5~MB~files.
Here, we can see a clear disadvantage for the horizontally scaled network.
Although this is apparent for the majority of circuits,
it becomes especially obvious for the long tail of circuits, the slowest~10\%.
Here, the transfer time increases by up to~25\%.

There are multiple possible influential factors for this effect.
First, we suppose that a resource allocation problem exists:
Tor balances traffic evenly across the network
by assigning the consensus weight of each relay based on its bandwidth.%
However, the circuit selection is carried out independently
by each client as a random selection process.
Consequently, optimal utilization of the network resources
is not guaranteed by cooperative allocation, but probabilistically.
We informally argue from a combinatorial perspective
that the horizontally scaled network offers more possible combinations for circuits,
making optimal choices less likely.
A second possible influence could be that the directory servers,
which distribute the consensus data to the clients,
do not profit from horizontal growth as much as from vertical growth.
Finding the specific root cause would require much deeper investigations
and are out of scope of this work.
However, the results demonstrate the merits of our proposed methodology,
showing that it can be used to generate novel, nontrivial insights about scaled Tor networks.

In total, we come to the conclusion that the way of scaling the Tor network
can have a significant influence on its performance.
While vertical scaling generally appears to retain the performance of the original Tor network,
horizontally scaled networks may lead to a considerable drop in performance.
We argue that it is important to study such effects
before growing the network in one or the other direction.
Our methodology for artificially scaling Tor consensuses can serve as an essential building block for this.

\section{Related Work}

Our work is in line with several other pieces of research
that focus on the scalability of the Tor network.
Improved scalability of the Tor network has been identified
as an important condition for further growth~\cite{DBLP:journals/csur/AlSabahG16}.
Consequently, many approaches for better scalability have been put forward~\cite{walking-onions,torsk,pir-tor}.
Up to now, however, the focus was mainly on improving or maintaining performance in larger networks.
In contrast, our work adds a methodology suitable to assess
\emph{the impact} of network growth---in terms of performance, security,
or any other aspect that can be evaluated on the basis of Tor consensuses.
The existing research contribution that comes closest to our work
is~\cite{once-never-enough}, introducing tornettools.
At first glance, tornettools does something similar to torsynth: It scales Tor consensuses.
However, there are several core differences.
Firstly, tornettools can only scale the network \emph{down} by sampling subnetworks.
This is a fundamentally different task because no relays have to be \enquote{invented}.
Moreover, tornettools was designed
to make it feasible to input existing snapshots of Tor to the shadow simulator.
In contrast to torsynth, it does not allow changing the shape of the Tor network,
\eg, based on bandwidth distribution or relay roles.
Also, tornettools does not take families into account.

As our work allows to generate Tor consensuses of scaled network instances,
it complements a variety of existing research that uses consensuses,
allowing them to be applied for scalability analysis as well.
One prime example is TorPS~\cite{users-get-routed}.
TorPS is a tool for simulating circuit selection Tor.
It has successfully been used for security research in Tor~\cite{users-get-routed,tempest}
as well as performance research~\cite{waterfilling}.
Such research can immediately benefit from our work
by feeding TorPS with consensuses generated by our torsynth.
Moreover, network simulation of Tor can benefit.
In the past, shadow has been used extensively for Tor research
focusing both on performance~\cite{kist,peerflow}
as well as security~\cite{point-break}.
Used in combination with tornettools that processes Tor consensuses,
these can be applied to the analysis of scaled towards networks as well.
The same is true for other research approaches
that rely on Tor consensuses in other ways~\cite{navigator,onions-in-the-crosshairs}.
With the ongoing and future evolution of the Tor network,
analysis of deployment strategies for scaled networks would be another use case~\cite{doepmann18deployingonions}.

\section{Conclusion}

In this work, we presented a methodology to utilize real-world network data
for exploring challenges of network growth in Tor.
Our implementation, torsynth, will enable researchers
to leverage their existing methods
for the analysis of hypothetical, scaled Tor networks,
as will be necessary to identify and cope with challenges of the future Tor.
Applying torsynth for Tor network simulations,
we show that massive horizontal network growth of Tor
may lead to suboptimal performance.
As future work, torsynth is envisioned to serve as a versatile
and extensible framework for working with synthetic consensuses.

\begin{acks}
This work has been partially funded by the
Deutsche Forschungsgemeinschaft
(DFG, German Research Foundation, TS 477/1-1).
\end{acks}

\onecolumn
\begin{multicols}{2} %
   \printbibliography
\end{multicols}

\end{document}